\documentclass[a4paper]{article}
\usepackage{a4wide}
\usepackage{hyperref}

\newcommand{\anchor}[1]{#1}
\newcommand{\toprule}{\hline}
\newcommand{\midrule}{\hline}
\newcommand{\bottomrule}{\hline}

\usepackage{enumitem}
\AtBeginDocument{%
	\providecommand\BibTeX{{%
		\normalfont B\kern-0.5em{\scshape i\kern-0.25em b}\kern-0.8em\TeX}}}

\begin{document}

\title{Going dark? Analysing the impact of end-to-end encryption on the outcome of Dutch criminal court cases}



\author{Pieter Hartel \and
        Rolf van Wegberg
}
\date{%
    \{pieter.hartel,r.s.vanwegberg\}@tudelft.nl\\[2ex]%
    \today
}

\maketitle

\begin{abstract}
Law enforcement agencies struggle with criminals using to end-to-end encryption (E2EE).
A recent policy paper states:
``while encryption is vital and privacy and cyber security must be protected, that should not come at the expense of wholly precluding law enforcement''.
The main argument is that E2EE hampers attribution and prosecution of criminals who rely on encrypted communication - ranging from drug syndicates to child sexual abuse material (CSAM) platforms.
This statement - in policy circles dubbed `going dark' - is not yet supported by empirical evidence.
That is why, in our work, we analyse public court data from the Netherlands to show to what extent law enforcement agencies and the public prosecution service are impacted by the use of E2EE in bringing cases to court and their outcome.
Our results show that Dutch courts appear to be as successful in convicting offenders who rely on E2EE as those who do not.
Our data does not permit us to draw conclusions on the effect of E2EE on criminal investigations.
\end{abstract}






\section{Introduction}

Every technology has a bright and a dark side~\cite{Felson2019}.
For example, GPS was designed to guide missiles to their targets~\cite{Enge1999}.
Civil use was a secondary objective, but now we would be lost without GPS-based navigation.
Mobile phones meet one of the most basic human needs: the ability to communicate.
But drug dealers and their customers also love their phones because they no longer have to meet in a dark alley to avoid the police.

End-to-end encryption (E2EE) is a system that allows mobile phone users to communicate with each other without anyone else eavesdropping.
So, the police cannot listen in either, even if they are authorized to tap the communication.
PGP was the first widely used implementation of E2EE~\cite{Zimmermann1996a}, and WhatsApp has been offering E2EE since April 2016 to over a billion users~\cite{Menezes2021}.
PGP has helped human rights organizations and journalists to communicate in hostile environments.
PGP has probably saved hundreds of lives in the Kosovo theatre \href{https://philzimmermann.com/EN/letters/}{\anchor{Letters from human rights groups}}.
But offenders use PGP phones~\cite{Rourke2020} to defeat lawful interception.
A PGP phone is a relatively expensive product on which not only PGP is installed, but from which also all non-essential hard and software have been removed~\cite{Europol2020}.

The content of the communication may be encrypted, but the location of the phones is not.
Every PGP phone has a regular phone number and uses the same mobile phone network as all other mobile phones.
Suppose PGP phone A sends a message to PGP phone B.
Then the encrypted message first goes to a cell tower near A, then via the network of the Telco to a cell tower near B, and finally from the cell tower to B.
The police can locate a PGP phone by asking the provider when and with which cell towers the phone was in contact.
Location information has been successfully used in several lawsuits to breach the anonymity of PGP phone users.
For example, a court judgment describes how cell tower data proved that the telephones of an offender and his co-offender travelled together from Eindhoven to Amsterdam, where both were stopped during a traffic control \href{https://uitspraken.rechtspraak.nl/inziendocument?id=ECLI:NL:RBAMS:2016:2835}{\anchor{ECLI:NL:RBAMS:2016:2835}}.
Law enforcement has other special powers to bring offenders to justice that we will discuss below.


So, all these technologies have two sides, but to what extent does the dark side have the upper hand?
We probably agree that navigation with GPS and communication with a mobile phone has so many advantages that we accept the disadvantages.
But do the advantages of E2EE also outweigh the disadvantages?

E2EE only works properly if it is correctly implemented in a trustworthy execution environment and if the private keys remain secret.
However, this is more easily said than done.

In recent law enforcement operations against \emph{criminal service providers} such as Phantom Secure, IronChat, Ennetcom, EncroChat, and Sky ECC, the police have managed to obtain messages - e.g., by infiltration - whereas the companies claimed that this should be impossible.
The police were legally allowed to take action against these criminal service providers since there was a well-founded suspicion that these companies provided services to criminals.
For example, Phantom Secure was a Canadian company that was infiltrated by FBI employees in 2018.
Recorded conversations with the Phantom Secure CEO led to a valid allegation that the company's modified Blackberry phones were used for drug trafficking~\cite{Europol2020}.
Offenders not only use PGP, but they also use WhatsApp.
For example:
``The fact that the offender sold these drugs came to light after four young adults became unwell from drugs they had bought after WhatsApp contact with a dealer'' \href{https://uitspraken.rechtspraak.nl/inziendocument?id=ECLI:NL:RBNNE:2018:5197}{\anchor{ECLI:NL:RBNNE:2018:5197}}.

Offenders use PGP and WhatsApp for different reasons.
Most users probably know that WhatsApp offers E2EE, but they do not seem to care about it~\cite{Abu-Salma2017}.
WhatsApp is a success because almost all the people you want to communicate with are already using it - i.e., the network effect.
WhatsApp is easy to use, free and even ad-free.
PGP phones on the other hand, are an expensive niche product.
The users buy such a device because the confidentiality of the messages they exchange with it is of vital importance to them.
Specialised companies sell PGP phones and service subscriptions at premium prices.
Offenders might use WhatsApp to communicate with clients or victims, but they might use a PGP-phone to communicate with co-offenders.

In the Netherlands, several Ennetcom court cases have now been concluded, and some of the court judgments have been made public as open data.
To gain insight into the impact of E2EE on the outcome of Dutch criminal court cases, we will analyse these and other relevant court judgments.

\section{Background and research questions}
In the Netherlands, law enforcement has a wide range of special powers at their disposal, as described in Article 126 of the \href{https://wetten.overheid.nl/BWBR0001903}{\anchor{Code of Criminal Procedure}}.
The application of these powers is subject to strict rules.
In particular, special powers may only be used for serious offences, and permission from the examining magistrate is required.
It should also be possible to check afterwards whether the powers have been used correctly.
These checks and balances are in place to ensure a fair trial.
Technical special powers that are often used in investigations where the offender tries to evade detection through technology are (1) reading out and analysing confiscated smartphones, (2) placing telephone or Internet taps, (3) obtaining cell tower data from a Telco to trace the location of a mobile phone, and (4) hacking the computer or another device of the offender.
There are other special powers, such as a subpoena for financial data, systematic observation, and systematic gathering of information, but we will not consider these here since they are not specifically designed to deal with technology such as E2EE.
We will describe in more detail below two often-used special powers that on the one hand suffer from encryption, but on the other hand provide useful data.

\paragraph{Phone data}
Most modern devices have encryption turned on by default.
This means, that data on seized devices can only be read out if the device owner supplies the passcode.
Law enforcement has several options to obtain phone data.
\begin{itemize}[noitemsep,topsep=0pt]
\item The owner may surrender the passcode to the police.
This should not be done under duress because, in most countries, as the offender should not be obliged to cooperate with his conviction (nemo tenetur)~\cite{Rourke2020}.
\item In some countries, the police may force one to give up a fingerprint to unlock a smartphone~\cite{Europol2020}.
\item In some cases, special tools can bypass the passcode.
For example, to crack the San Bernardino terrorist's iPhone 5C, the FBI had to pay more than \$ 1M to a specialist company~\cite{Cate2018}.
\item With the permission of the examining magistrate, the police may install key logger malware on a smartphone.
The key logger reports the passcode without the suspect knowing~\cite{Brown2020}.
\end{itemize}

\paragraph{Server data}
Lawful interception allows authorised law enforcement agencies to obtain communication network data from individual subscribers.
The signalling and network management information will be clear text, for example, IP addresses.
The contents of the data can be encrypted, for example, when HTTPS or E2EE is used.
In almost all implementations of E2EE, devices communicate with each other through a server.
Law enforcement has several options to obtain server data:
\begin{itemize}[noitemsep,topsep=0pt]
\item If the server contains a bug, an exploit can be used to tap the communication.
This has happened to \href{https://www.wired.com/story/whatsapp-hack-phone-call-voip-buffer-overflow/}{\anchor{WhatsApp}}.
\item If the administrators of the server make mistakes, the server can be hacked.
This has happened to \href{https://www.vice.com/en/article/3aza95/how-police-took-over-encrochat-hacked}{\anchor{EncroChat}}.
\item If the administrators of the server are issued a subpoena by the court to hand over data from specific customers, they will have to comply.
This has happened to \href{https://www.wired.com/2007/11/encrypted-e-mai/}{\anchor{HushMail}}.
\item If law enforcement can pose as a reseller of handsets, they can insert a backdoor into the handset before delivering them to the customer.
This has allegedly happened to \href{https://www.vice.com/en/article/epd3km/sky-ecc-hacked-fake-app}{\anchor{Sky ECC}}.
\item The police can also take the servers down and arrest the owners.
This has happened to \href{https://www.vice.com/en/article/a34b7b/phantom-secure-sinaloa-drug-cartel-encrypted-blackberry}{\anchor{Phantom secure}}.
\end{itemize}

\paragraph{Research questions}
The law ensures that an offender is only convicted if all evidence is legally obtained and conclusive.
Suppose, that the content of a message from an offender is encrypted.
The court may still be able to see to whom the offender has sent the message, but the court does not learn the content of the message.
Then, the message could be legal evidence, but the court will probably deem it inconclusive.
Also, assume that there is no other evidence, just the encrypted message.
Then, all cases where the offender has used E2EE will lack conclusive evidence and are either not brought to court or lead to an acquittal by the court.
This is a hypothetical situation, as there may be enough other evidence to convict the offender, for example, location data.
It does not matter whether the offender has used a PGP phone or WhatsApp, because in both cases, the phone must communicate regularly with a cell tower.
The Telco therefore knows the location of the phone in question.
And, with the location data obtained from the Telco, the court may decide that the evidence is conclusive.
Because E2EE may reduce the number of options that law enforcement has to collect legal and convincing evidence, our first research question is:
\textit{To what extent does law enforcement use its special powers when offenders resort to E2EE? (RQ1)}

Cases for which the police cannot obtain sufficient evidence are normally not tried in court.
We have made inquiries at the \href{https://www.forensicinstitute.nl}{\anchor{Netherlands Forensic Institute}}, but unfortunately, no public data or statistics are available on these types of cases.
Our analysis is, therefore, limited to cases brought to the courts.
Because acquittal can be a consequence of the use of E2EE, our second research question is:
\textit{To what extent are offenders using E2EE acquitted? (RQ2)}

A court judgment is a decision about the offender.
However, a judgment also contains information about other persons involved in an investigation, such as co-offenders but also unknown persons with a criminal role.
If unknown persons appear more often in E2EE investigations, this could be an indication that E2EE hinders the work of the police.
E2EE use can lead to a lack of evidence, but it is usually not the only cause.
For example: ``Because entering the offender's home was unlawful, the objects found may not be used as evidence.
Because there is no other evidence, in this case, the offender is acquitted'' \href{https://uitspraken.rechtspraak.nl/inziendocument?id=ECLI:NL:RBAMS:2020:4884}{\anchor{ECLI:NL:RBAMS:2020:4884}}.
We could investigate the occurrence of unknown persons in our three groups, which would shed more light on the problem caused by E2EE.
Instead, we decided to zoom in onto the PGP group -- as this group represents a worst case for law enforcement -- and focus on a potential solution: the capture by law enforcement of server or phone data from criminal service providers. 
We pose as the third research question:
\textit{To what extent do unknown persons occur in investigations using data from criminal service providers. (RQ3)}

\section{Method}
In six years (2015 - 2020), the Dutch district courts published 25,366 anonymized court judgments on \href{https://rechtspraak.nl}{\anchor{rechtspraak.nl}}.
This represents about 5\% of the total number of court judgments in that period.
The courts publish all judgments with a crime against life, where the maximum sentence is at least four years, or when the court expects interest from the public.
Therefore, judgments of the most serious crimes are likely to be included in the published data set.

Offenders and the police are engaged in an on-going battle.
As soon as one wins, the other tries to nullify that lead.
E2EE gives the offender a head start, and the question is to what extent the special powers of the police can cope.
We will therefore construct a comparison group of judgments in which the police used their special powers, but in which the offender did not use E2EE.
These judgments form a baseline for judgments in which the offender \emph{has} used E2EE.

To answer RQ1, we will compare the group of judgments where the offender has used E2EE to the comparison group.
To answer RQ2, we will compare the conviction rates of the three groups.
To analyse the court judgments, we define three variables as follows:
\begin{itemize}[noitemsep,topsep=0pt]
\item The first variable \textit{special power} encodes the technical special powers used by law enforcement in reaction to the offender using E2EE.
\item The second variable \textit{decision} encodes whether the offender is convicted or acquitted.
\item The third variable \textit{technology} encodes whether the offender used PGP, WhatsApp, or neither (comparison).
A judgment with both WhatsApp and PGP is considered a PGP judgment; the three groups are therefore independent.
\end{itemize}

To answer RQ3, we will identify judgments that mention unknown persons in a criminal role as a proxy for potential co-offenders who have not been tried due to lack of evidence.
We assume, that such unknown persons have stayed anonymous, because they used PGP.
The analysis is therefore worst-case because every potential offender who has not been prosecuted is blamed on the use of PGP.
The unit of assessment for RQ3 is therefore the investigation, rather than the judgment as for RQ1 and RQ2.
An investigation is defined as the set of judgments originating from the same police investigation.
To analyse the police investigations, we define two variables: \textit{data} and \textit{others}.
The first variable \textit{data} encodes whether or not law enforcement has used data from a criminal service provider.
There are two possibilities:
\begin{itemize}[noitemsep,topsep=0pt]
\item Server data has been obtained from criminal service providers, such as Ennetcom, EncroChat, PGP-safe, Sky-ECC, and IronChat, or from police operations Onymous and Bayonet.
\item The PGP phone of the offender has been read out or his PGP keys were seized.
\end{itemize}
The second variable \textit{others} encodes whether unknown persons played a criminal role in the investigation.
This can be stated in many ways, for example: ``The offender is a career criminal of the worst kind who lives in circles where liquidation orders are given and received'' \href{https://uitspraken.rechtspraak.nl/inziendocument?id=ECLI:NL:RBAMS:2017:5136}{\anchor{ECLI:NL:RBAMS:2017:5136}}.
We have also looked for judgments stating that the case against one of the co-offenders has been dropped.
However, this does not occur in the PGP judgments.

\paragraph{Descriptive statistics}
A total of 6,619 relevant court judgments were available for analysis.
This is about 1.5\% of the total number of criminal judgments processed by the Dutch district courts in the given 6-year period.
In 439 judgments PGP was used, WhatsApp was used in 2,390 judgments, and the comparison group consists of 3,790 judgments.
The groups are unbalanced, which weakens some of the statistical analysis.
We sampled 20\% of the WhatsApp group and 12\% of the comparison group (both uniform and at random).
This gave us a WhatsApp group of 437 judgments and a comparison group of 469 judgments, in total N=1,345.

Of the 1,345 judgments, 25.5\% were drugs-related, and 26.6\% were violence-related.
These percentages are higher than the national averages of 9.7\% and 9.2\% respectively~\cite[Table 6.2 and 6.12]{Meijer2021} because the courts mainly publish judgments of serious crimes.

The offender is female in 7.9\% of judgments.
The average age of the offender at the time of the court judgment is 36.2 (SD = 12.5) years.
Of the offenders, 37.9\% are first-time offenders, and 31.5\% are repeat offenders.
These demographics are consistent with the demographics of the whole population of Dutch criminal offenders convicted for serious crime~\cite{Wingerden2016}.

Of the 1,345 judgments, 80.0\% have resulted in incarceration, including involuntary commitment, imprisonment, and military detention.
The average length of incarceration is 42.7 (SD = 50.0) months, which is more than 10 times the national average of 4 months~\cite[Table 6.11]{Meijer2021}, again because of the focus on serious crime.
Community service represents 5.2\%, acquittal 6.3\%, and a fine 3.0\%.
The remaining 4.7\% of the judgments are procedural, such as an extradition request.

The police have used their technical special powers as follows:
In 68.0\% of judgments, a phone or Internet connection was tapped (offenders with a PGP-phone may also have a regular phone).
In 26.9\% of judgments, a seized mobile phone was read out.
In 10.9\% of judgments, a phone was located by requesting cell tower data.
The Dutch police have hacked into the offender's systems eight times in 2019, just after passing the relevant law that made this possible.
However, none of those judgments are public (yet), so that we have no data on police hacks.

The 439 PGP judgments are the result of 196 criminal investigations.
In the majority of these (83.3\%), law enforcement used server or phone data from criminal service providers.

\begin{table}[hhhh]
\caption{Contingency table of court judgments using specific \textit{technology} (left) versus \textit{offence type} (top)
($\chi^2(10)=350.48, p<0.001$, Cramer's V$=0.36, p<0.001$, $\alpha=0.01$.)}
\label{tab:offence_type}
\begin{tabular}{l *{6}{r} | r r }
\toprule
Judgments       &Property       &Violent        &Public         &Drug           &Weapon         &Other          & \multicolumn{2}{c}{Total} \\
                &offence        &offence        &order          &offence        &offence        &criminal       & row           & column \\
                &               &               &offence        &               &               &offence        &               &     \\
\midrule
  WhatsApp      & 119           & 182           & 39            & 41            & 11            & 45            & 437           & \\
                & 27.2\%        & 41.6\%        & 8.9\%         & 9.4\%         & 2.5\%         & 10.3\%        & 100\%         & 32.5\% \\
  PGP           & 30            & 60            & 50            & 233           & 12            & 54            & 439           & \\
                & 6.8\%         & 13.7\%        & 11.4\%        & 53.1\%        & 2.7\%         & 12.3\%        & 100\%         & 32.6\% \\
  Comparison    & 122           & 116           & 41            & 69            & 12            & 109           & 469           & \\
                & 26.0\%        & 24.7\%        & 8.7\%         & 14.7\%        & 2.6\%         & 23.2\%        & 100\%         & 34.9\% \\
\midrule
  Total         & 271           & 358           & 130           & 343           & 35            & 208           & 1,345         & \\
                & 20.1\%        & 26.6\%        & 9.7\%         & 25.5\%        & 2.6\%         & 15.5\%        & 100.0\%       & 100.0\% \\
\bottomrule
\end{tabular}
\end{table}


\section{Results}
Table \ref{tab:offence_type} tabulates the crime rates for the main offence types defined by Statistics Netherlands \href{https://www.cbs.nl/nl-nl/onze-diensten/methoden/classificaties/misdrijven/standaardclassificatie-misdrijven-2010}{\anchor{cbs.nl}}.
\textit{Other criminal offence} includes offences not covered by any of the other categories, for example, road traffic offences, and environmental crime.
Sometimes procedural judgments are not tied to a specific offence, for instance, extraditions.
The ``Other'' column also accounts for these procedural judgments.
An offender may commit more than one crime, but we have counted only the offence with the most severe maximum sentence.
A $\chi^2$ test of association between \textit{technology} and \textit{offence type} was found to be statistically significant (see caption).
This means, that the difference in crime rates between the three groups is unlikely to exist due to chance.
For example, offenders using WhatsApp commit mostly violent crime (41.6\%), whereas PGP offenders mostly commit drugs-related offences (53.1\%).

\begin{table}[hhhh]
\caption{Contingency table of court judgments using specific \textit{technology} (left) versus \textit{special power} used by law enforcement (top)
($\chi^2(6)=336.31, p<0.001$, Cramer's V$=0.35, p<0.001$, $\alpha=0.01$,
and for the table without the first column with a zero cell count:
$\chi^2(4)=94.73, p<0.001$, Cramer's V$=0.30, p<0.001$, $\alpha=0.01$).}

\label{tab:special_power}
\begin{tabular}{l *{4}{r} | r r }
\toprule
Judgments       &No             &Tapped         &Readout        &Located        &\multicolumn{2}{c}{Total} \\
                &special        &only           &w/ or w/o      &w/ or w/o      & row           & column \\
                &power          &               &tapped         &tapped, readout&               & \\
\midrule
  WhatsApp      & 177           & 153           & 85            & 22            & 437           & \\
                & 40.5\%        & 35.0\%        & 19.5\%        & 5.0\%         & 100.0\%       & 32.5\% \\
  PGP           & 87            & 145           & 123           & 84            & 439           & \\
                & 19.8\%        & 33.0\%        & 28.0\%        & 19.1\%        & 100.0\%       & 32.6\% \\
  Comparison    & 0             & 338           & 91            & 40            & 469           & \\
                & 0.0\%         & 72.1\%        & 19.4\%        & 8.5\%         & 100.0\%       & 34.9\% \\
\midrule
  Total         & 264           & 636           & 299           & 146           & 1,345          & \\
                & 19.6\%        & 47.3\%        & 22.2\%        & 10.9\%        & 100.0\%       & 100.0\% \\
\bottomrule
\end{tabular}
\end{table}



Table~\ref{tab:special_power} shows the relationship between the variables \textit{technology} and \textit{special power}.
The police prefer the tap (47.3\%) to reading out phones (22.2\%) and gathering cell tower data (10.9\%).
By the construction of the comparison group, special powers were used in all comparison judgments.
A $\chi^2$ test of association between \textit{technology} and \textit{special power} was found to be statistically significant (see caption).
This means, that the difference in the use of special powers between the three groups is unlikely to exist due to chance.
For example, the police use special powers more for PGP (100-19.8=80.2\%) than for WhatsApp (100-40.5=59.5\%) judgments.

\begin{table}
\caption{Contingency table of court judgments using specific \textit{technology} (left) versus \textit{decision} (top)
($\chi^2(2)=3.09, p=0.213$, Cramer's V=$0.05, p=0.21$, $\alpha=0.01$).}
\label{tab:decision}
\begin{tabular}{l *{2}{r} | r r }
\toprule
Judgments       &Convicted      &Acquitted      & \multicolumn{2}{c}{Total} \\
                &               &               & row           & column \\
\midrule
  WhatsApp      & 405           & 22            & 427           & \\
                & 94.8\%        & 5.2\%         & 100.0\%       & 33.3\% \\
  PGP           & 397           & 28            & 425           & \\
                & 93.4\%        & 6.6\%         & 100.0\%       & 33.2\% \\
  Comparison    & 395           & 35            & 430           & \\
                & 91.9\%        & 8.1\%         & 100.0\%       & 33.5\% \\
\midrule
  Total         & 1,197          & 85            & 1,282          & \\
                & 93.4\%        & 6.6\%         & 100.0\%       & 100.0\% \\
\bottomrule
\end{tabular}
\end{table}



Table~\ref{tab:decision} shows the relationship between the variables \textit{technology} and \textit{decision}.
To focus on the differences between conviction and acquittal, we have omitted the procedural judgments; hence the total number is 1,282 instead of 1,345.
In all three groups, the vast majority of offenders is convicted.
A $\chi^2$ test of association did not reveal a significant difference between the conviction rates of the three groups (see caption).
This means, that there is no evidence in our data that the outcome of a trial depends on whether the offender used PGP, WhatsApp or neither.

\begin{table}
\caption{Contingency table of investigations using specific \textit{data} (left) versus the criminal involvement of \textit{others} (top)
($\chi^2(1)=4.426, p=0.035$, Cramer's V$=0.15, p<0.001$, $\alpha=0.01$). }
\label{tab:others}
\begin{tabular}{l *{2}{r} | r r }
\toprule
Investigation   & No others     & With others   & \multicolumn{2}{c}{Total} \\
                &               &               & row          & column \\
\midrule
Without server/ & 55            &  69           & 124 \\
phone data      & 44.4\%        &  55.6\%       & 100.0\%      & 63.3\% \\
With server/    & 21            &  51           &  72 \\
phone data      & 29.2\%        &  70.8\%       & 100.0\%      & 36.7\% \\
\midrule
Total           & 76            & 120           & 196          & \\
                & 38.8\%        &  61.2\%       & 100.0\%      & 100.0\% \\
\bottomrule
\end{tabular}
\end{table}

Table table~\ref{tab:others} shows the relationship between the variables \textit{data} and \textit{others}.
The 439 judgments constitute 196 investigations.
In 36.7\% of the investigations, criminal service provider data was used, and in 61.2\% unknown persons were involved.
A $\chi^2$ test showed that there is no significant relationship between the variables.
This means, that there is no evidence in our data that the availability of server or phone data influences the number of investigations with unknown others.

\section{Discussion}
\paragraph{Research questions}
The answer to RQ1 is that law enforcement uses more special powers in cases where offenders use PGP than where they use WhatsApp.
This places a burden on law enforcement and ultimately on the taxpayer.
However, law enforcement does not use all its special powers, and it does not use special powers for all investigations either.
This can be explained by assuming that in a PGP investigation, a mere Internet tap would not provide enough data to create conclusive evidence.
This suggests, that law enforcement has more special powers than it currently uses.

The answer to RQ2 is that there is no evidence in our dataset that the conviction rate of offenders who use EE2E differs from the conviction rate of offenders who do not use EE2E.
This means, that our data shows no evidence that the outcome of court decisions is influenced by E2EE.


The answer to RQ3 is that the data available to us shows no difference in the extent to which unknown persons are criminally involved in investigations with, or without, data from criminal service providers.
This means, that our data shows no evidence that capturing data from criminal service providers influences the number of unknown persons with a criminal role.
However, our data does show that the majority of police investigations into PGP cases (83.3\%) use criminal service provider data or phone read outs.

\paragraph{Public-policy debate}
We provide some observations as a contribution to the public-policy debate~\cite{Hewson2021}.

Some courts seem to hint towards legislative action against criminal use of E2EE, as evidenced by phrases from court judgments such as:
``This crypto phone belongs to the accused and is of such a nature that its uncontrolled possession is contrary to the law or the public interest.'' \href{https://uitspraken.rechtspraak.nl/inziendocument?id=ECLI:NL:RBZWB:2020:1216}{\anchor{ECLI:NL:RBZWB:2020:1216}}.
What the courts have probably not considered is whether controlling possession is feasible.
If the legislator restricts the use of EE2E, the authorities would have to verify that all service providers duly implement the restrictions.
We think that this would be a heavier burden on governments (and on the taxpayer) than the status quo.

Next to the burden of additional police costs to work around E2EE, there are other interests too~\cite{Veen2020}.
For example, national security agencies are unlikely to use backdoor encryption because of the risk of the key to the back door ending up in the wrong hands.
And confidentiality is crucial for national security agencies.
Also, the commercial use of E2EE with a back door would probably not be viable because of the risk that a competitor would get hold of the keys.
This means, that many legitimate users of E2EE will find alternative means of secure communication that law enforcement will not be able to tap, thus aggravating the problem for law enforcement rather than ameliorating it.


If E2EE is weakened - or in essence, broken - by policies that demand a back door, a supra-national infrastructure is needed to manage those backdoors.
Every nation-state will need to access backdoors to prosecute its nationals, including states on the EU sanctions list.
We believe, that this is a recipe for disaster.
Banning E2EE will simply force terrorists, drug dealers, and paedophile rings to use alternative technologies.
Well-funded offenders are already starting to develop their own encryption platforms~\href{https://www.vice.com/en/article/wjwbmm/inside-the-phone-company-secretly-run-by-drug-traffickers}{\anchor{MPC}}.
Initially, such tools will have issues, but over time they will get better and will create an obstacle to law enforcement.

Law enforcement currently does an excellent job of taking down criminal service providers like EncroChat.
Recent law enforcement operations against these companies show that there are sufficient opportunities to monitor them and to act upon information that shows their involvement in illegal activity.
Our recommendation is not to build a back door into every application of E2EE, but to keep a watchful eye on relevant, criminal service providers.

\paragraph{Limitations}
Our analysis is focused on PGP and WhatsApp, as only nine judgments mention Signal and two mention Telegram~\cite{Albrecht2021}.
The most importantly limitation is that we do not know when special powers have proven insufficient for law enforcement to build a case because such information is confidential.
Instead, we have used acquittal by the courts as an indication of inconclusive evidence.
The data we have used stems from the Dutch government and is not necessarily representative of other countries.
The data also only represents about 1.5\% of all criminal judgments in the Netherlands.

\section{Conclusions}
The criminal justice system is often described as a funnel that inputs many more crime reports than that it outputs convicted offenders~\cite{Felson2019}.
Few crime reports lead to a police investigation, and even fewer investigations lead to a court case.
The court judgments that we have been able to analyse make the convictions transparent but the rest of the funnel remains opaque.
However, by searching for criminal roles played by unknown others, we have tried to see beyond convictions, and into the investigations.

The information position of technology companies and governments today is superior to that of the nineties due to surveillance from online and offline sources.
Encryption is one of the few technologies available to law-abiding citizens, corporations, and national security agencies that protect privacy.
Yet, criminals (mis)use that same technology.

We have shown that the courts in a democracy with sufficient checks and balances can do its work without legislation that breaks encryption.
We cannot make a similar conclusion for law enforcement as our data is inconclusive on this point.
One way to gain insight into this problem is by examining police files and interviewing police detectives.
This we suggest as future work.

\section*{Acknowledgment}
Conversations with Phil Zimmermann have been a great source of inspiration for this work.
We thank Roel Wieringa and the anonymous referees for their comments on the paper.
The research complies with ethical standards because all data that has been analysed is open data that has been made public to be analysed.

\bibliographystyle{plain}

\end{document}